\documentclass[11pt]{amsart}
\usepackage[slovene, english]{babel}
\usepackage{graphicx,amsmath}
\usepackage{wrapfig}
\usepackage{caption}
\usepackage{tikz,lmodern}
\usepackage{array}
\usepackage{empheq}
\usepackage{commath}
\usepackage{hyperref}
\usepackage{subfig}
\usepackage{float}
\usepackage[format=hang,justification=raggedright,
singlelinecheck=false,margin=10pt,font=small,labelfont=bf]{caption}
\usepackage[fitting,skins,theorems]{tcolorbox}
\hypersetup{
	    colorlinks,
	    linkcolor={red!50!black},
	    citecolor={blue!50!black},
	    urlcolor={red!20!black}
}

\definecolor{mathcolor}{HTML}{1A3CA4}
\everymath{\color{mathcolor}}

\title{Precession of the elastic pendulum on the rotating Earth}
\author{B. Jur\v{c}i\v{c} - Zlobec}
\thanks{Faculty of Electrical Engineering, University of Ljubljana}
\def\Oomega{\mathit{\Omega}}

\lccode`"`"

\begin{document}
\selectlanguage{english} 
\begin{abstract}
We present a numerical solution of the nonlinear differential
equation for a pendulum with an elastic string on the rotating Earth, for different values
of string stiffness at different geographic latitudes.

We are looking for the critical stiffness of the spring at which the transition  
between $24$-hour precession of the oscillating plane of a pendulum with an ideal elastic string 
and the precession of a Foucault’s pendulum with an ideal stiff string occurs. 
The precession of a Foucault’s pendulum is $24/\sin\lambda$, where $\lambda$ is the latitude. 

The transition between the precessions is 
gradual. However, we observed another dynamic that is related
with a loss of the phase of the pendulum oscillation
due to circular motion of the pendulum. 
This work represents a more detailed investigation of this dynamics. 

We found that the loss in phase in Ljubljana $(46.05^\circ N, 14.5^\circ E)$ occurs at 
coefficient $\delta = 0.06$. In this case, the ratio between a relaxed and a loaded string 
is $l_0/l_e = 0.11$. The time $\tau$ is equal to $12.125$\,h, 
and the period of Foucault's pendulum revolution is equal to $33.40$\,h.
(See Figure \ref{fig:cont}.)
\\[2mm]
\end{abstract}
\maketitle
\section{Introduction}

In \cite{Stanovnik}, A. Stanovnik discusses the oscillation of a pendulum with an ideal 
elastic string on a rotating Earth. 
The ideally elastic string is a Hook's spring that has zero length at relaxation.

In this case, the transversal and longitudinal frequencies of the pendulum are the same. 
In this article we vary the stiffness of the spring and observe the oscillations depending on latitude. 
Our observations focus on 
the precession of the projection of the pendulum's motion in the plane tangential to Earth at a specific point.

Initially, the primary goal of the study was to observe the transition between the 24-hour 
precession period of a pendulum with an ideal elastic string and the $24/\sin\lambda$-hour 
period of Foucault's pendulum, which has an ideally rigid (inextensible) string, 
depending on the latitude and string stiffness.
The transition between one regime and the other is gradual; the oscillations in general are not periodic.

If we look at the oscillation of a pendulum with an ideally flexible string we notice that the direction 
of oscillation remains unchanged relative to the stellar background. 
On the $45^\circ$ parallel 
the oscillation of a pendulum swinging in the north-south direction will gradually change from a transversal to a longitudinal oscillation
over 12 hours. In longitudinal oscillation, the precession phase of the oscillation plane is lost. 

\begin{wrapfigure}{l}{0.5\textwidth}
\includegraphics[height=0.5\textwidth]{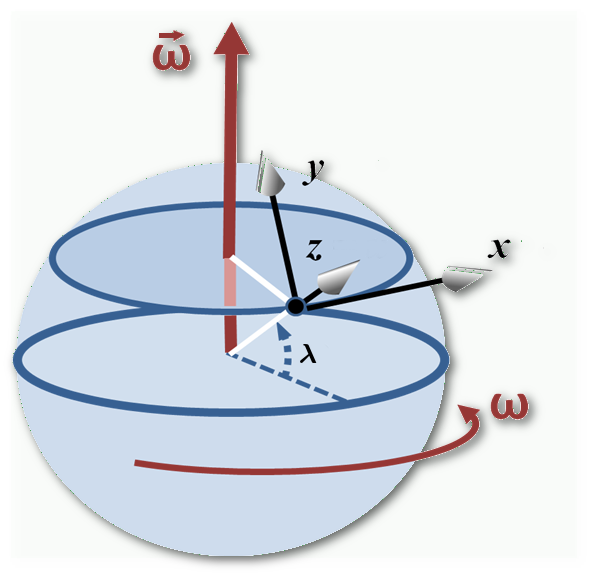}
\caption{Coordinates system}\label{fig:koo}
\end{wrapfigure}

Precession phase loss is well defined, so we
focus on observing the loss of precession phase at different geographic
latitudes for different spring stiffness. On a given geographical
latitude, the pendulum swings in the north-south direction changing the stiffness of the spring
until at some point the projection of the pendulum motion becomes circular.

At latitude of $45^\circ$, the stiffness at that point must be zero,
and the pendulum oscillates only in the longitudinal direction (the projection of the motion is a point).
At latitudes different from $45^\circ$, the initial trajectory of the pendulum bob in space is a stretched ellipse, 
which then gradually opens and the eccentricity decreases until, at a certain stiffness and latitude, 
the projection of the trajectory becomes circular. Under the same initial conditions, the diameter of the circle depends on the latitude.

In addition to stiffness, we are also interested in the time required at a given 
latitude for the projection motion of the pendulum to become circular.
We will call the stiffness at which the projection of the pendulum forms a circle
the critical stiffness for a given latitude.
The stiffness will 
be measured by the relative difference between the longitudinal and transversal angular 
frequencies of the pendulum's oscillation, relative to the frequency of Earth's rotation.

It turns out that for stiffness defined in this way, critical stiffness of a pendulum is characteristic 
for a given latitude, regardless of the length of the pendulum, gravitational acceleration, 
and period of rotation of the planet, as long as we can ignore the centrifugal reaction force acceleration.
Therefore, there is no need to restrict to Earth anymore, and we will from now on refer to a planet to emphasize this independence.

As previously stated, the motion of a pendulum with critical stiffness of the string is generally not periodic. 
Our main interest lies in identifying those latitudes at which the pendulum oscillation 
becomes periodic at critical stiffness. The search for such resonant latitudes will be the focus of our study in what follows.

\section{Equation for an elastic pendulum}
The time it takes for one rotation (period) of a planet around its axis (one day) is denoted by $T$. 
The angular frequency of the planet is therefore $\omega=2\pi/T$. 
Geographical latitude will be denoted by $\lambda$. 
We will observe the oscillation of a bob suspended on an elastic string at the given latitude 
taking into consideration the rotation of the planet around its axis.

As defined above, the stiffness or elasticity of the string on which the pendulum bob is suspended is  
measured by the relative difference between the longitudinal and transversal angular 
frequencies of the pendulum's oscillation, normalized by the angular frequency of the planet's 
rotation around its axis. After giving the pendulum an initial push in the north direction, 
the bob will initially describe a stretched ellipse in space. 
We will observe the precession of the projection of the semi-major axis of the ellipse 
onto the $(x,y)$ plane, and the precession angle will be denoted by $\phi$.

The length of the unloaded pendulum string will be denoted by $l_0$, and the 
length of the loaded string in equilibrium by $l_e$. 
The mass of the bob is denoted by $m$, and gravitational acceleration by $g$. 

Longitudinal angular frequency of the pendulum is equal to 
$\omega_s=\sqrt{k/m}$, where $k$ is the spring stiffness coefficient.
The transversal angular frequency of the pendulum is $\omega_p=\sqrt{g/l_e}$, 
where $l_e$ is the length of the loaded string in equilibrium. 
The spring stiffness coefficient $k$ and longitudinal angular 
frequency can be expressed in terms of $l_0$, $l_e$, and $g$ as:
\begin{equation}\label{equ:ll}
mg=k(l_e-l_0),\quad k= \frac{mg}{l_e-l_0},\quad
\omega_s=\sqrt{\frac{g}{l_e-l_0}}.
\end{equation}

The string stiffness coefficient $\kappa$ 
is expressed as the ratio of the longitudinal and transversal angular 
frequencies squared
\begin{equation}\label{equ:kappa}
	\kappa=\frac{\omega_s^2-\omega_p^2}{\omega_s^2}=\frac{l_0}{l_e}.
\end{equation}
The relative stiffness of an ideally flexible string is equal to 0 ($l_0=0$), 
while the relative stiffness of an inextensible string is equal to 1. 
In our case, instead of using the stiffness coefficient, we will use the 
relative difference between the longitudinal and transversal angular 
frequencies as a measure of stiffness.

\begin{equation}\label{equ:delta}
         \delta=\frac{\omega_s-\omega_p}{\omega_p}.
\end{equation}

The differential equation was solved in the local coordinate system that moves 
together with the pendulum. We set up the system as shown in image \ref{fig:koo}. 
We chose the starting point of the coordinate system at the mount point of the 
pendulum so that the position vector of the equilibrium position of the bob is 
at $(0,0,-l_e)$. We denote the position vector of the pendulum bob by $\vec{r}$ 
and the unit vector in the direction of the $z$ axis, as is usual, by $\vec{k}=(0,0,1)$. 
The absolute value of a vector is denoted by the same letter without an arrow above it, 
$r=\abs{\vec{r}}$. The gravitational acceleration vector in this coordinate system is $\vec{g}=g(0,0,-1)$.
The angular frequency vector of the planet is equal to
$\vec{\omega}=\omega\:(0,\cos\lambda,\sin\lambda)$.

The Lagrangian of a system is the difference between its kinetic and potential energies.
\begin{equation}\label{equ:lag}
L= \frac{m}{2}\left(\frac{d\vec{r}}{dt}+
\vec{\omega}\times(\vec{R}+\vec{r})\right)^2
         - mg\,\vec{k}\cdot\vec{r}
         -\frac{m}{2}\frac{g}{l_e-l_0}(r-l_0)^2.
\end{equation}
The first part is the kinetic energy. The speed of the pendulum bob is 
composed of two contributions: the velocity in the local coordinate system, 
to which the speed due to the rotation of the earth is added. 
The potential energy consists the potential energy in 
Earth's gravitational field to which the potential energy due to the 
stretching of the string is added.

The vector $\vec{R}$ represents the position of the pendulum string attachment point in 
the coordinate system, which has its origin at the center of the planet. 
It contributes to a uniform centrifugal reaction, which can be taken into account 
by correcting the gravitational force (effective gravitational force: 
$m\vec{g}_{eff}=m(\vec{g}-\omega\times(\omega\times\vec{R}))$. 
We have decided not to consider this system's force for now.

To write the Lagrangian $L$ in dimensionless quantities, 
we first divide it by the mass $m$ and introduce dimensionless quantities 
$\rho=r/l_e$ and $\tau=t\omega_p$. The angular frequency is then measured 
in the transversal angular frequency of the pendulum string. 
We denote the quotient $\omega/\omega_p$ by $\Oomega$, while the quotient 
$\omega_s/\omega_p$ is expressed from the equation \eqref{equ:delta}, 
which gives $\omega_s/\omega_p=1+\delta$. 
The Lagrangian expressed in normalized quantities then becomes

\begin{equation}
         L=\frac{1}{2}\left(\frac{d\vec{\rho}}{d\tau}+
         \vec{\Oomega}\times\vec{\rho}\right)^2-
         \vec{k}\cdot\vec{\rho}-
         \frac{1}{2}\left((1+\delta)\rho-
         \frac{\delta(2+\delta)}{1+\delta}\right)^2.
\end{equation}

The conserved quantity, the Hamiltonian, is
\begin{equation}\label{equ:energy}
         H = \frac{\partial L}{\partial\dot{\vec{\rho}}}\,\dot{\vec{\rho}}-L
         =\frac{1}{2}\,\dot{\vec{\rho}}{\,^2}-
         \frac{1}{2}
         (\vec{\Oomega}\times\vec{\rho})^2+
         \vec{k}\cdot\vec{\rho}+
         \frac{1}{2}\left((1+\delta)\rho-\frac{\delta(2+\delta)}{1+\delta}\right)^2,
\end{equation}
where $\dot{\vec{\rho}}=d\vec{\rho}/d\tau$.

The moment $\vec{p}$ is expressed as
\begin{equation}\label{equ:moment}
         \vec{p} = \frac{\partial L}{\partial \dot{\vec{\rho}}} =
         \dot{\vec{\rho}} + \vec{\Oomega}\times\vec{\rho}.
\end{equation}
To analyze the dynamics of the system, we need to express the Hamiltonian in 
canonical variables $\vec{p}$ and $\vec{q}=\vec{\rho}$
\begin{equation}
         H =\frac{1}{2}\,\vec{p}{\,^2} -
         \vec{p}\cdot(\vec{\Oomega}\times\vec{q})+
         \vec{k}\cdot\vec{q}+
         \frac{1}{2}\left((1+\delta)\,q-\frac{\delta\,(2+\delta)}{1+\delta}\right)^2,
\end{equation}
and write Hamilton's system of equations
\begin{equation}
\tcboxmath[colframe=black!50!blue,boxrule=0.4pt]{
       \dot{\vec{q}} = \frac{\partial H}{\partial\vec{p}} =
       \vec{p} - \vec{\Oomega}\times\vec{q},
}
\end{equation}

\begin{equation}\label{equ:hamilton}
\tcboxmath[colframe=black!50!blue,boxrule=0.4pt]{
      \dot{\vec{p}} = -\frac{\partial H}{\partial\vec{q}} = 
      -\vec{\Oomega}\times\vec{p}  
      - \vec{k}
      -(1+\delta)^2\vec{q} + \delta(2+\delta)\,\frac{\vec{q}}{q}.
}
\end{equation}
To obtain the differential equation of motion, 
we insert the Lagrangian into the Euler-Lagrange equation 
%which relates the derivatives of the Lagrangian with respect to the 
%generalized coordinates and time to the derivatives of the Lagrangian 
%with respect to the corresponding momenta and time
\begin{equation}
\frac{d}{d\tau}\left(\frac{\partial L}{\partial\dot{\vec{\rho}}}\right)-
         \frac{\partial L}
         {\partial\vec{\rho}}=0.
\end{equation}
The equation of oscillation is therefore 
\begin{equation}\label{equ:def}
\tcboxmath[colframe=black!50!blue,boxrule=0.4pt]{
\frac{d^2\vec{\rho}}{d\tau^2} + \vec{k} +
\vec{\rho}(1+\delta)^2 -
\delta(2+\delta)\frac{\vec{\rho}}{\rho} +
2\,\vec{\Oomega}\times\frac{d\vec{\rho}}{d\tau} -
\vec{\Oomega}\times(\vec{\rho}\times\vec{\Oomega})=0.
}
\end{equation}
It can be seen from equation \eqref{equ:def} that the behavior 
of the system depends only on the relative difference between 
the angular frequencies of the trans\-ver\-sal and longitudinal oscillations of the pendulum, 
denoted by $\delta$, and the vectors of the relative angular frequency 
of the planet's rotation, denoted by $\vec{\Oomega}=\vec{\omega}/\omega_p$. 
We observe that the equation in \eqref{equ:def} differs from the usual 
equation of the Foucault's pendulum due to the additional terms containing the coefficient $\delta$.

\section{Pendulum with an ideal elastic string}
As previously mentioned, for a pendulum with an ideal elastic string, 
the longitudinal and transversal angular frequencies are equal, 
and the stiffness coefficient is 0.
The pendulum bob will be given an initial speed in the direction of the y-axis, towards the north. 
As a result, the bob will trace out a stretched ellipse while maintaining the 
orientation of its semi-major axis relative to the stellar background.

The precession of the projection of the semi-major axis onto the plane $(x,y)$ will be observed 
The precession period of the pendulum in this case is equal to the rotation period of the planet. 
The precession of the semi-major axis does not necessarily make a complete turn. In this case, 
the precession period is the time when the projection of the semi-major axis of the ellipse returns 
to its initial position in the plane $(x,y)$, regardless of whether it makes a full turn or not (comare \cite{Stanovnik}).

\begin{figure}[H]
\centering
\subfloat[\tiny $\lambda=30^\circ$ and $60^\circ$ dashed, and around $45^\circ$ full]{
\includegraphics[width=0.48\textwidth]{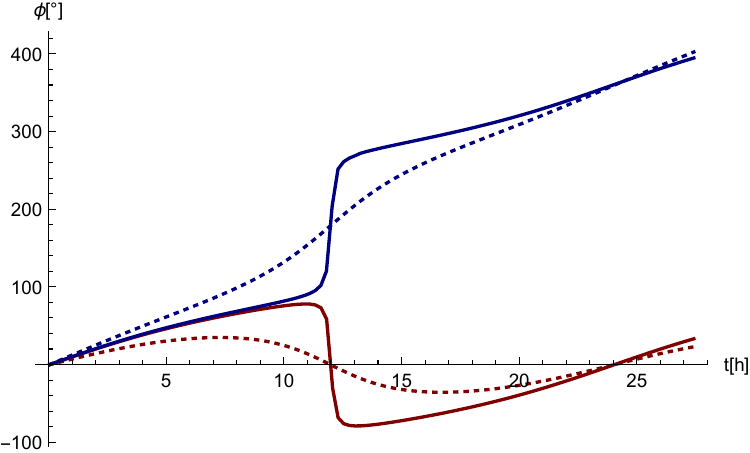}\label{fig:angle1}}
\subfloat[\tiny Comparison with Foucault's precession at $45^\circ$]{
\includegraphics[width=0.48\textwidth]{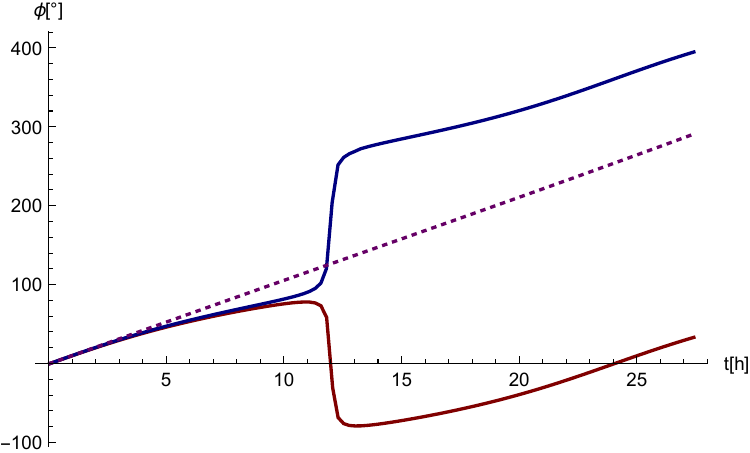}\label{fig:angle2}}
\caption[Precession]{Precession of a pendulum with an ideal elastic string}
\label{fig:angle}
\end{figure}
\noindent
Initially the angular velocity of precession is equal to the precession of  
Foucault's pendulum, that is $\mp\omega\sin\lambda$. 
The negative sign applies in the northern hemisphere, and the positive in the southern 
(see Figure \ref{fig:angle2}). At latitudes below the forty-fifth parallel, 
$0^\circ < \lambda < 45^\circ$, the pendulum does not make a full turn, 
while for $\lambda > 45^\circ$ it makes a full turn. 
The forty-fifth parallel is the boundary between the two different behaviors, 
as shown in Figure \ref{fig:angle1}. Below this parallel, the precession angle oscillates 
between the values
\begin{equation}\label{equ:osc}
         \varphi(\lambda)=\pm\left(
         \arctan\frac{\sqrt{\cos(2 \lambda )}}{\sin(\lambda )}+
         \frac{\pi }{2}\right),\quad 0\le \lambda \le \frac{\pi}{4}.
\end{equation}
The amplitude $\varphi$ approaches $\pi/2$ as we approach the $45^\circ$ 
parallel from below. At the $45^\circ$ parallel, the large semi-axis of the ellipse 
is vertical to the surface of the planet at the half period of the planet's 
rotation $T/2$. The pendulum oscillates only in the vertical direction. 
When the direction of oscillation passes the zenith point, the phase of the 
projection onto the plane $(x,y)$ suddenly changes by an angle of $\pm\pi$. 
The direction of the change, positive or negative, is not determined at the $45^\circ$ parallel. 
Slightly below this parallel, the phase changes for a short time by $-\pi$, 
and above it, the phase changes by $\pi$.

The pendulum precession completes a full turn above the $45^\circ$ parallel. 
The left figure on Figure \ref{fig:angle}, displays the time variation of 
the precession angle $\phi$ for an ideal elastic pendulum at latitudes $30^\circ$ 
and $60^\circ$ shown by dashed lines, and in the vicinity of $\lambda=45^\circ$ 
by a solid line. The time dependence is indicated in blue above and in red below the $45^\circ$ parallel.

\section{Solving the Differential Oscillation Equation}
The equation \eqref{equ:def} was solved using the {\tt Mathematica} package using the numerical solver for differential equations {\tt NDSolve[]}.

Initially, the pendulum was in equilibrium.
By a northward push the pendulum was given its initial velocity.
In our calculations the angular frequency $\omega_p$ was 
100 times higher than the angular frequency $\Oomega$, specifically
$\omega_p=1$ and $\Oomega=0.01$.

Time values were converted into hours so that 
one rotation of the planet around its axis lasted 24 hours.

\section{Changing the Stiffness of a Pendulum String}
Initially our goal was to find the relative difference 
between longitudinal and transverse angular frequencies for a given geographical latitude, 
such that the projection of pendulum motion at the same point would become circular.

We selected latitudes $25^\circ$ and $54^\circ$ to demonstrate the precession of 
the semi-major axis of the pendulum projection ellipse, as affected by the stiffness of the string. 
At a certain value $\delta$, the loss of phase due to circular motion occurs, 
resembling the loss of phase of a pendulum with an ideal flexible string at the $45^\circ$ parallel. 
This is illustrated in Figure \ref{fig:angle}.

\begin{figure}[H]
\centering
\includegraphics[width=0.48\textwidth]{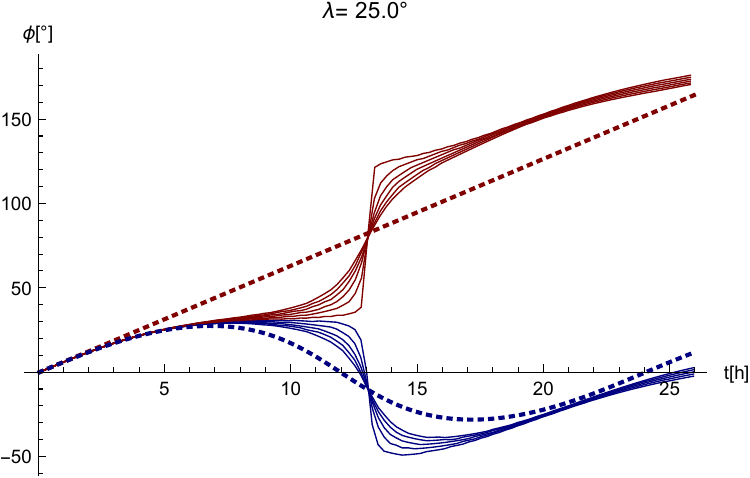}\hfill
\includegraphics[width=0.48\textwidth]{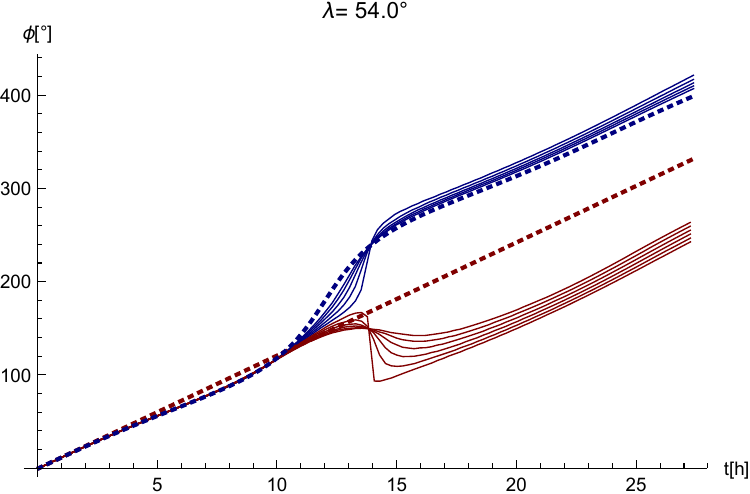}
\caption{Precession angle $\phi$ and time
below and above the $45^\circ$ parallel.}\label{fig:cont}
\end{figure}

The value of $\delta$ where this oocurs will be referred to as the critical stiffness that depends on the latitude. 
Figure \ref{fig:cont} illustrates the time dependence of the precession angle, with the angle measured in 
degrees on the y-axis and time in hours on the x-axis (with a period of $T=24$ hours). 
We selected a value of $\delta$ near the critical value for the latitude in question.

\begin{figure}[H]
\centering
\includegraphics[width=0.34\textwidth]{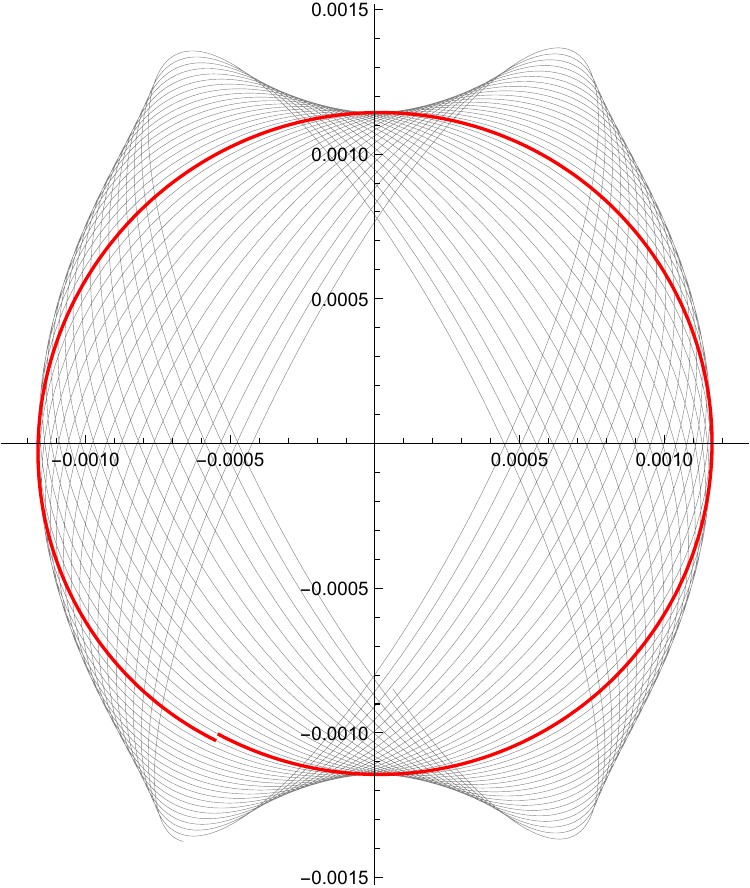}
\hspace{1.5cm}
\includegraphics[width=0.34\textwidth]{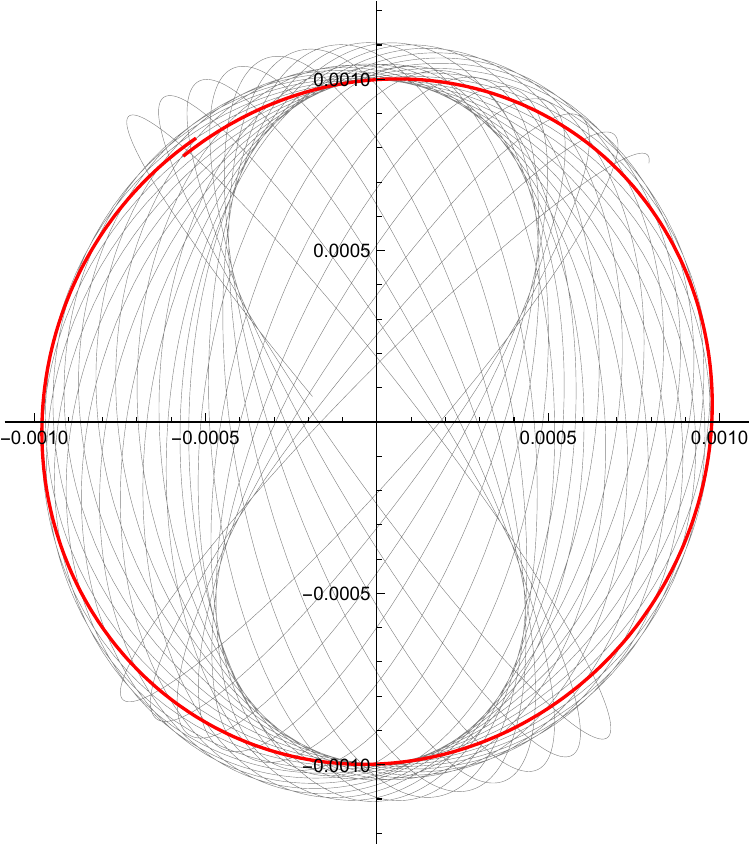}
\caption{The transition over the circular motion at 
$\lambda=20^\circ$ and $\lambda=60^\circ$.}\label{fig:detail}
\end{figure}

A dashed line shows the precession of the pendulum plane with an ideal elastic string (blue) 
and of Foucault's pendulum (red). The solid lines indicate the precession of the pendulum plane for 
values of $\delta$ slightly below (blue) and slightly above (red) the critical value for the given latitude. 
The graph depicting the precession angle as a function of time starts to diverge from the graph for an ideal 
elastic pendulum until it passes the critical value (causing phase loss), 
and then approaches Foucault's precession from above for latitudes below the 
$45^\circ$ parallel and from below for latitudes above it.
The graph of the projection of the bob's trajectory for the critical value $\delta$
reveals what happens at the moment when the loss of phase occurs and the projection of the bob's motion
becomes circular.  
\begin{figure}[H]
\begin{center}
\includegraphics[angle=90,width=0.48\textwidth]{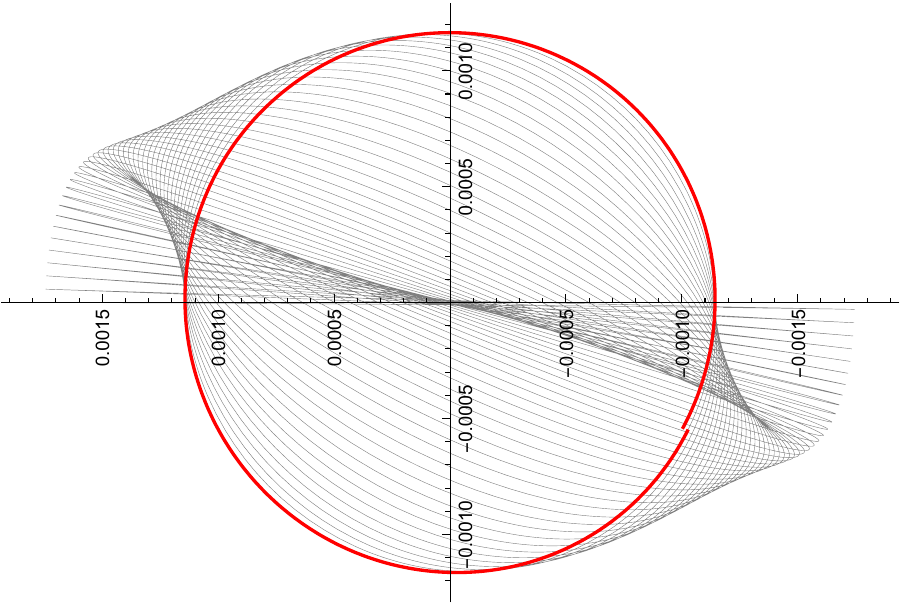}
\includegraphics[angle=90,width=0.48\textwidth]{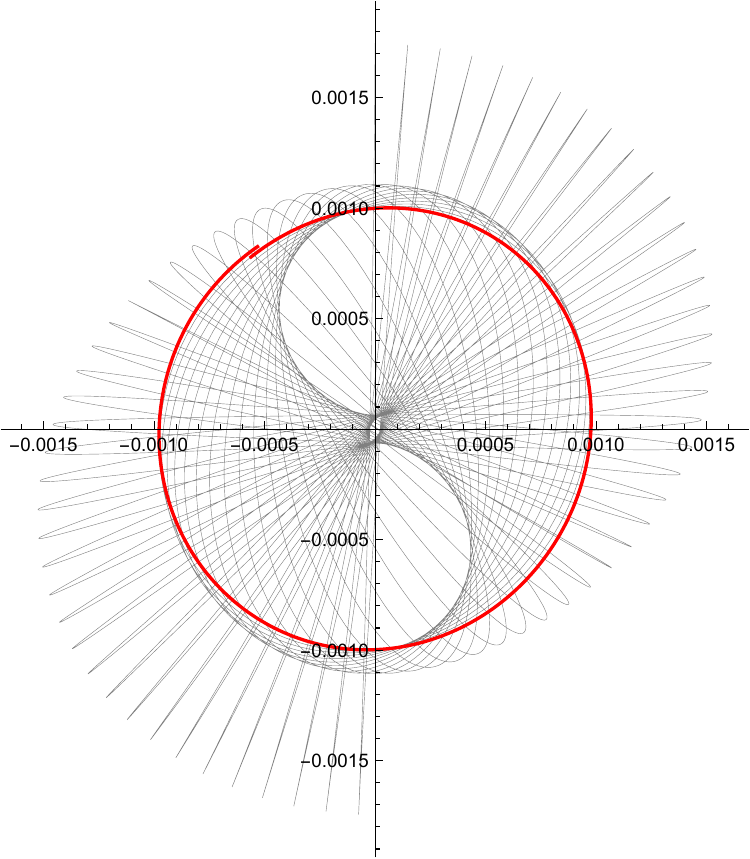}
\caption{Projection of the trajectory of the pendulum bob onto the plane $(x,y)$ for
$\lambda=20^\circ$ and $\lambda=60^\circ$.}
\label{fig:prec}
\end{center}
\end{figure}
The figure \ref{fig:detail} shows the change in eccentricity
projections of the trajectory of the pendulum bob at the transition over circular motion for parallels $20^\circ$ and
$60^\circ$.

\section{Critical stiffness and the time of phase loss}
We assume that the angular frequency of the planet is $1$ so time  will be measured in radians. 
One day is equivalent to $2\pi$ radians. Our objective is to observe the time it takes for the 
pendulum motion projection to become circular after swinging the pendulum initially in the north-south direction. 

Given that the deviations from the equilibrium position of the pendulum are small, 
and its period is much smaller than the rotation period of the planet 
the critical value of the coefficient $\delta$ is independent of the pendulum's 
length and is a characteristic of the latitude.
Our first step is to determine the critical value of the coefficient $\delta(\lambda)$ for each latitude $\lambda$. 

The time at which the pendulum's phase is lost at a given latitude $\lambda$ will be denoted by $\tau(\lambda)$.
Oscillation is not always periodic; the pendulum may not return to its initial position after time $2\tau$. 
However, for certain geographical latitudes, the oscillation becomes periodic. These latitudes will be referred to as  
\emph{resonant latitudes} and will be of interest in the following discussion.

The time $\tau(\lambda)$ when phase loss occurs increases with distance from the forty-fifth parallel. 
As we approach the poles and the equator it grows towards infinity. 

The critical value of $\delta$ approaches $1$ as we approach the poles and grows towards infinity as we approach the equator. 
Therefore, in the vicinity of the poles, the critical stiffness of the pendulum occurs when the longitudinal 
angular frequency is twice the transverse frequency.

\begin{figure}[H]
\includegraphics[width=0.49\textwidth]{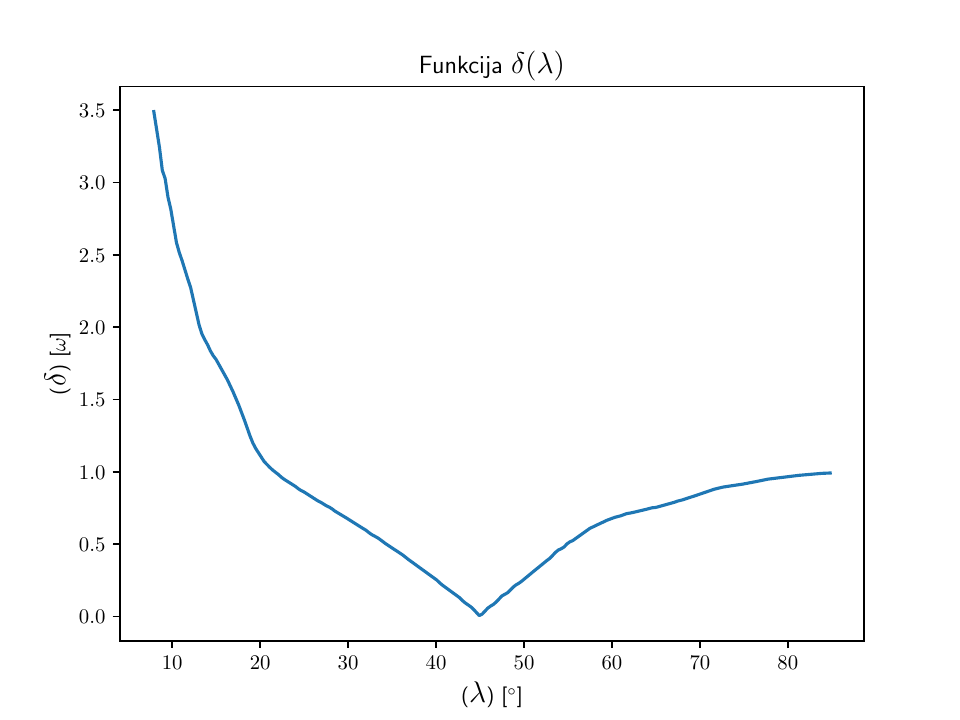}
\includegraphics[width=0.49\textwidth]{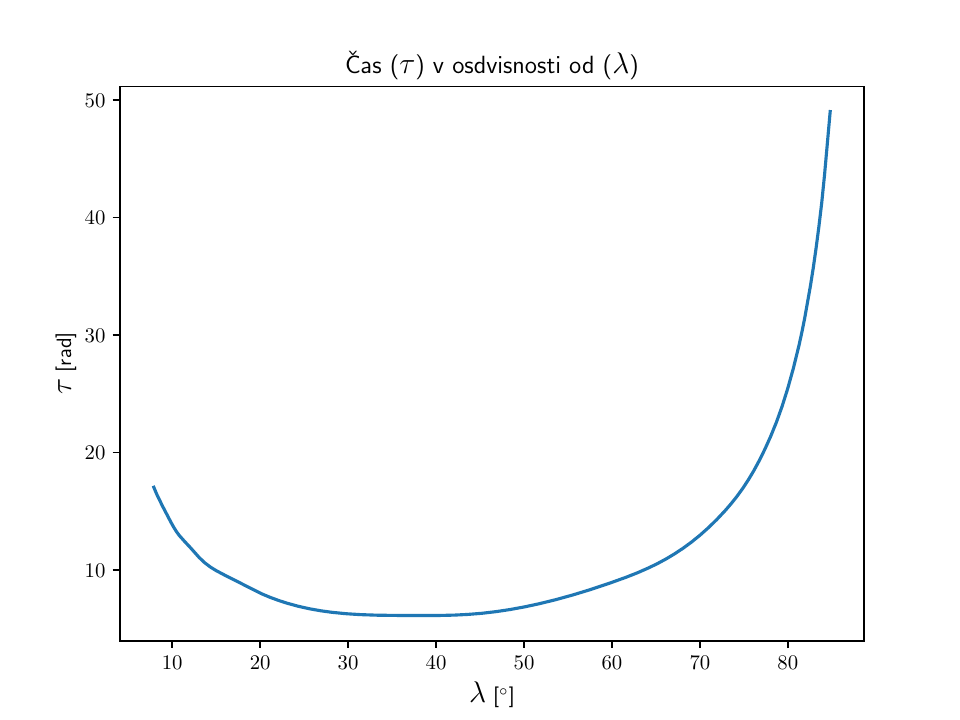}
\caption{The critical values of $\delta$ and $\tau$
depending on latitude $\lambda$.}\label{fig:delta-time}
\end{figure}

The value of $\delta$ becomes $1$ below the $45^\circ$ parallel at latitude $\lambda=21.5^\circ$. 
Between latitudes of $21.5^\circ$ and $90^\circ$, the critical value of $\delta$ decreases 
from $1$ to $0$ at the forty-fifth parallel and increases towards $1$ as we approach the north pole. 
Below $21.5^\circ$, the critical value of $\delta$ grows towards infinity as we approach the equator. 
Figure \ref{fig:delta-time} shows the dependencies of the critical value, $\delta(\lambda)$, 
and time, $\tau(\lambda)$, on $\lambda$.

\section{Deviation projection from rest position}
We will observe the changes of the maximum and minimum deviation of the projection 
of the pendulum's motion from the rest position. Initially, when the trajectory has the 
shape of a stretched ellipse, the deviation from the rest position changes from a 
value near 0 to a maximum value in one swing.

The difference between the maximum and minimum deviation from the rest position of the pendulum 
projection in the vicinity of the critical stiffness becomes smaller and smaller until it reaches zero, 
when the projection of the pendulum becomes circular, and then increases again.

\begin{figure}[H]
\includegraphics[width=0.495\textwidth]{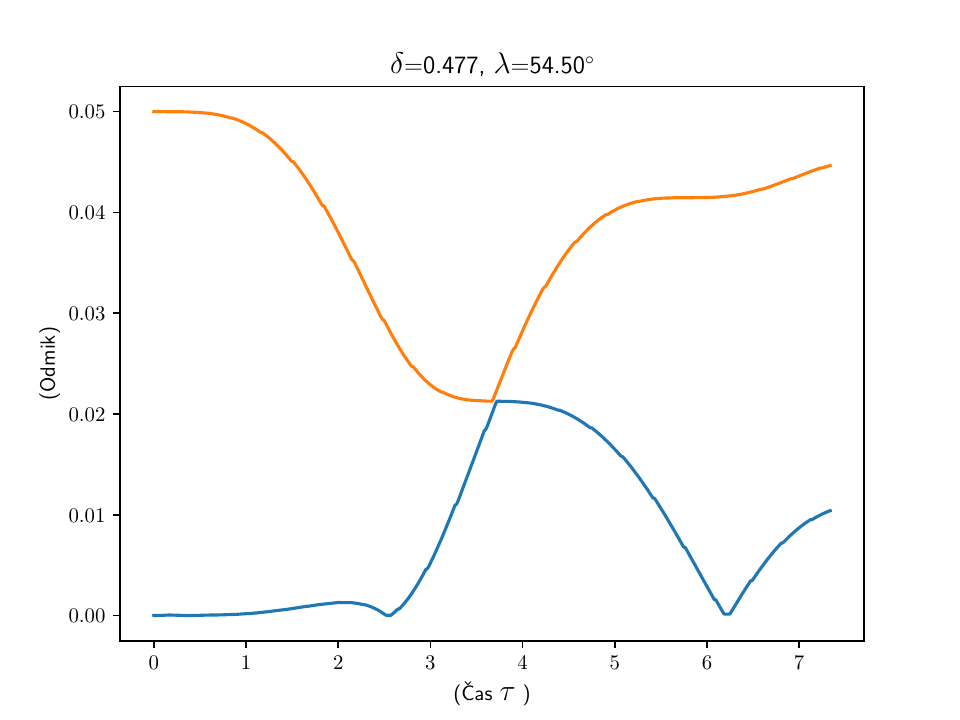}
\includegraphics[width=0.495\textwidth]{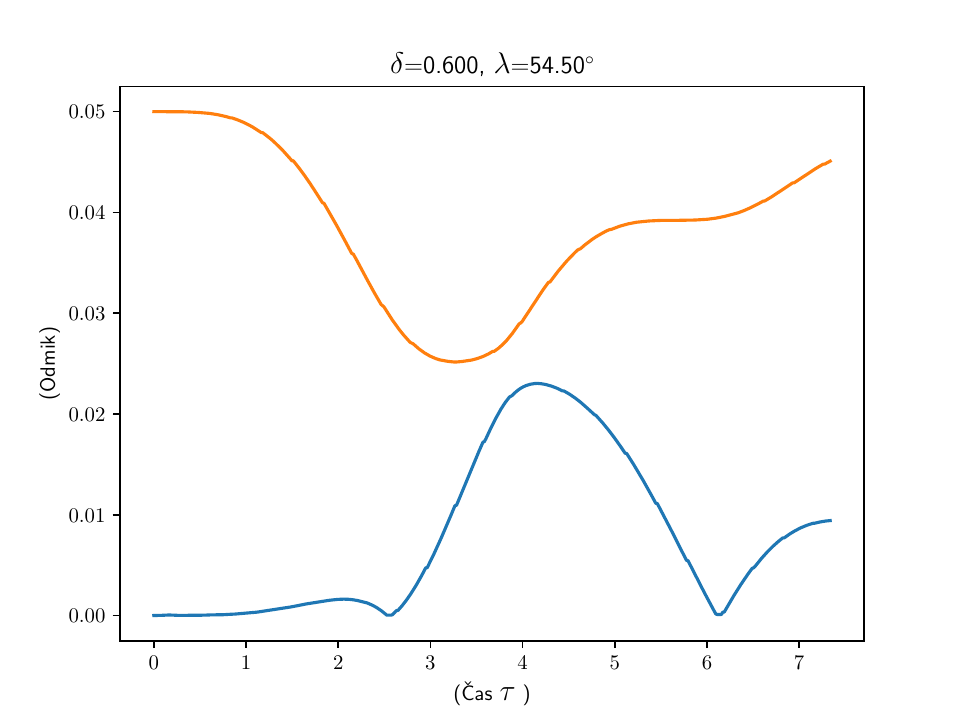}
\caption{Maximum and minimum value of pendulum projection deviation,
in the critical and supercritical value of stiffness, $\lambda=54.5^\circ$.}
\label{fig:moment}
\end{figure}

On the left of Figure \ref{fig:moment}, the bounds for the changes are shown for the maximum and minimum 
deviation of the projection of the pendulum's motion at the critical stiffness of the string, 
while on the right is the case with supercritical string stiffness. In the left picture 
the difference between the maximum and minimal deviation of the projection motion is close to zero in the middle. 
The time when this happens is $\tau$.
\begin{figure}[H]
         \includegraphics[width=0.495\textwidth]{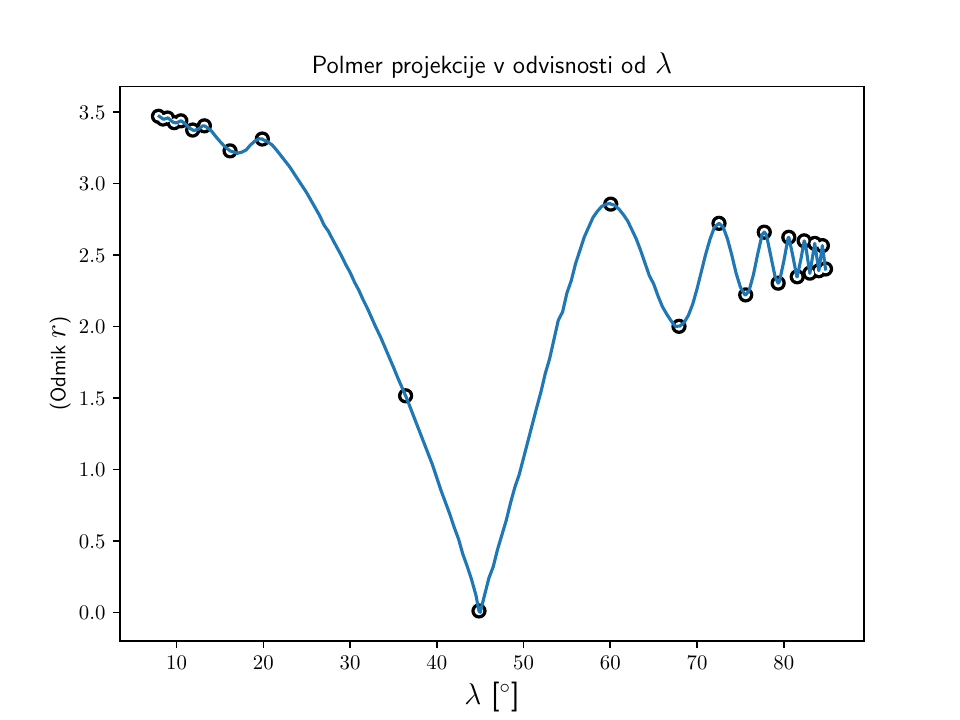}
         \includegraphics[width=0.495\textwidth]{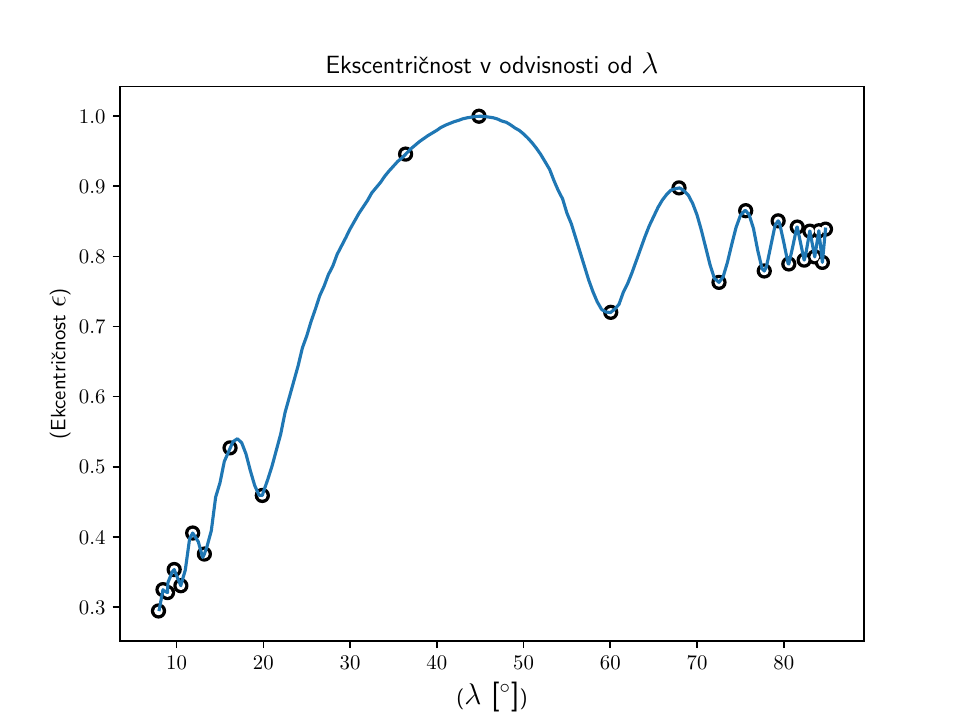}
\caption{Changing the projection radius and eccentricity of the trajectory.}\label{fig:radius}
\end{figure}
How does the projection radius and eccentricity of the spatial ellipse of the bob's 
trajectory change when the projection of the bob trajectory makes a circle, depending on latitude? 
The graph in Figure \ref{fig:radius} shows this dependency.

The oscillations of the radius and eccentricity curve attract attention. 
As we will see below, at latitudes of the local extrema, 
the oscillation of the pendulum is periodic with a period of $2\tau$.
We will call those latitudes as a \emph{resonant latitudes}.

Let us consider how the eccentricity of the spatial ellipse trajectory of the bob change within one period at 
(resonant) latitudes where the oscillation is periodic.
At latitudes below $45^\circ$, the eccentricity of the ellipse described by the bob trajectory 
at the moment $\tau$ is the smallest. In latitudes above $45^\circ$, however, 
the picture is more complicated. Here we notice that an additional oscillation is 
superimposed on the basic oscillation. 
The number of peaks of the superimposed oscillation is the order of the 
resonant latitude, counted from the parallel $45^\circ$.
The following figure shows two examples: one below the $45^\circ$ parallel and the other above it.

\begin{figure}[H]
\includegraphics[width=0.495\textwidth]{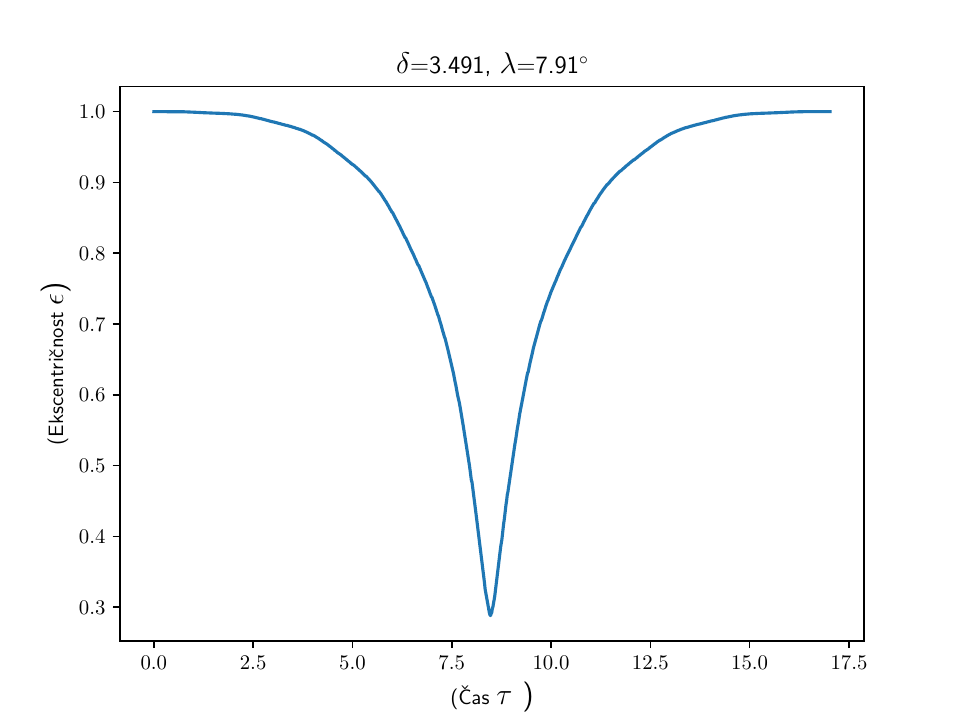}
\includegraphics[width=0.495\textwidth]{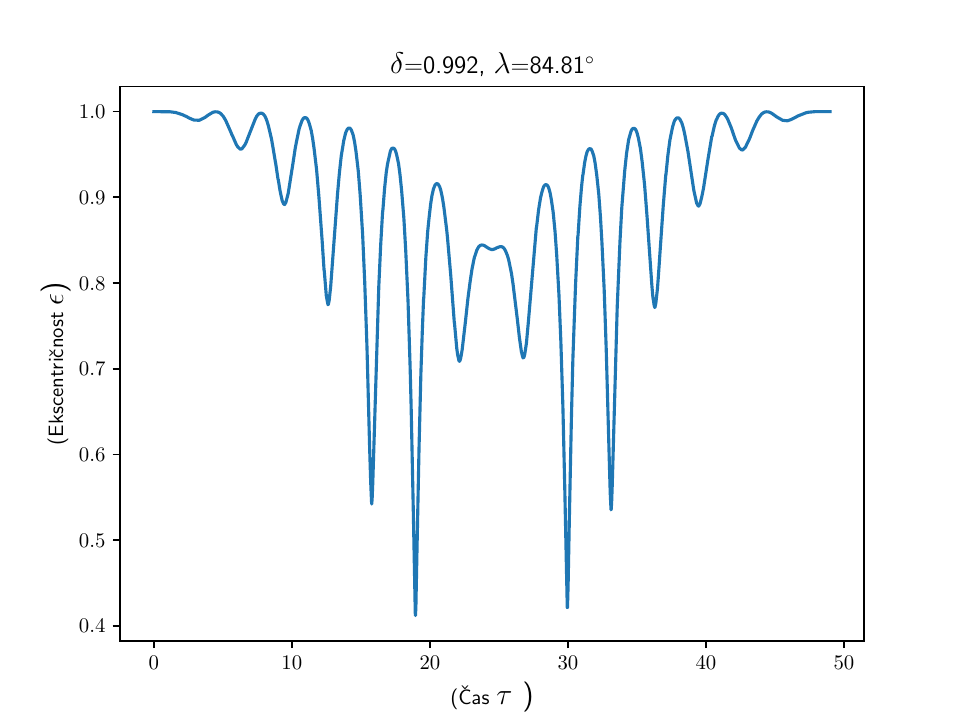}
\caption{Changing eccentricity, left $7.91^\circ$ 
	and right $84.81^\circ$}\label{fig:exc}
\end{figure}

\section{Resonant Latitudes}
As we have seen, oscillation of the pendulum for 
critical values of string stiffness is in general not periodic. However, at certain latitudes, 
called resonant latitudes, where the two curves in Figure \ref{fig:radius} reach local extrema, 
the oscillation of a pendulum with a critical string stiffness becomes periodic.

If we observe the extreme deviations of the longitudinal oscillations of the pendulum within one 
period at the resonant latitudes with critical stiffness we see that the basic periodic oscillation 
is modulated by an additional frequency that is superimposed on top of the fundamental oscillation. 
The number of oscillations of the imposed oscillation within the time interval $2\tau$ increases 
as we move away from the $45^\circ$ parallel.

The number of oscillations between two adjacent local extrema of the function in Figure \ref{fig:radius} differs by one. 
The number is even in local maxima, while it is odd in local minima.

\begin{figure}[H]
\includegraphics[width=0.32\textwidth]{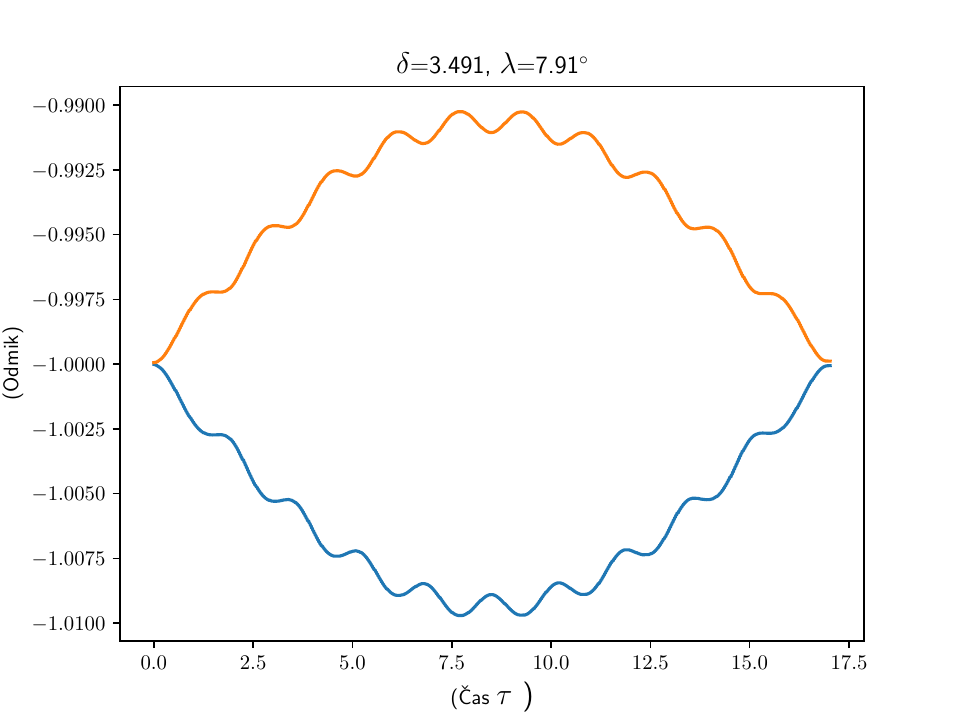}
\includegraphics[width=0.32\textwidth]{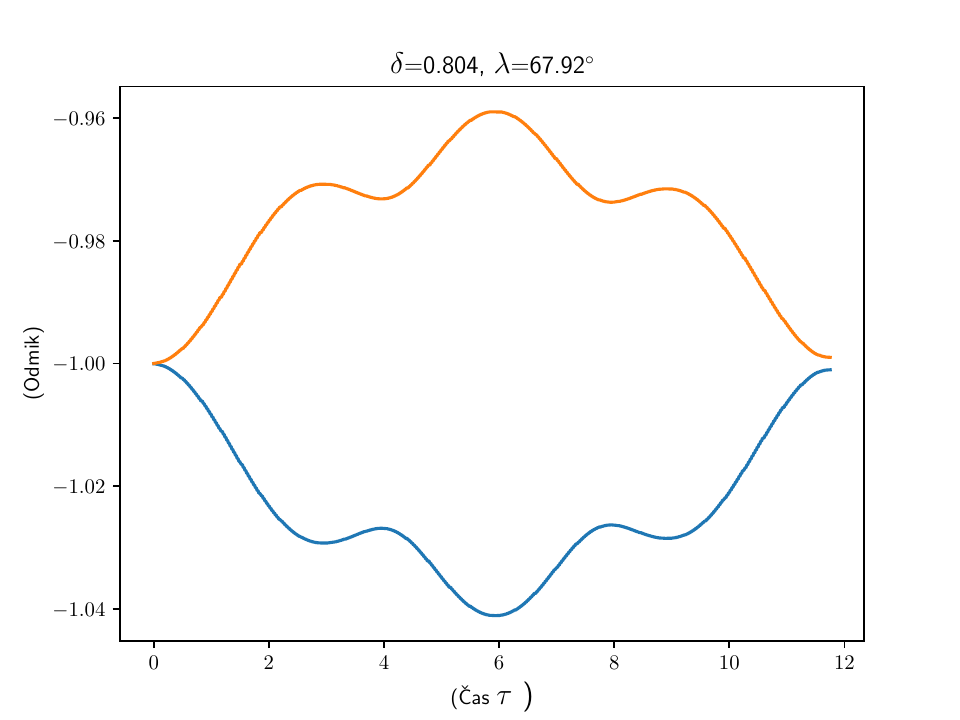}
\includegraphics[width=0.32\textwidth]{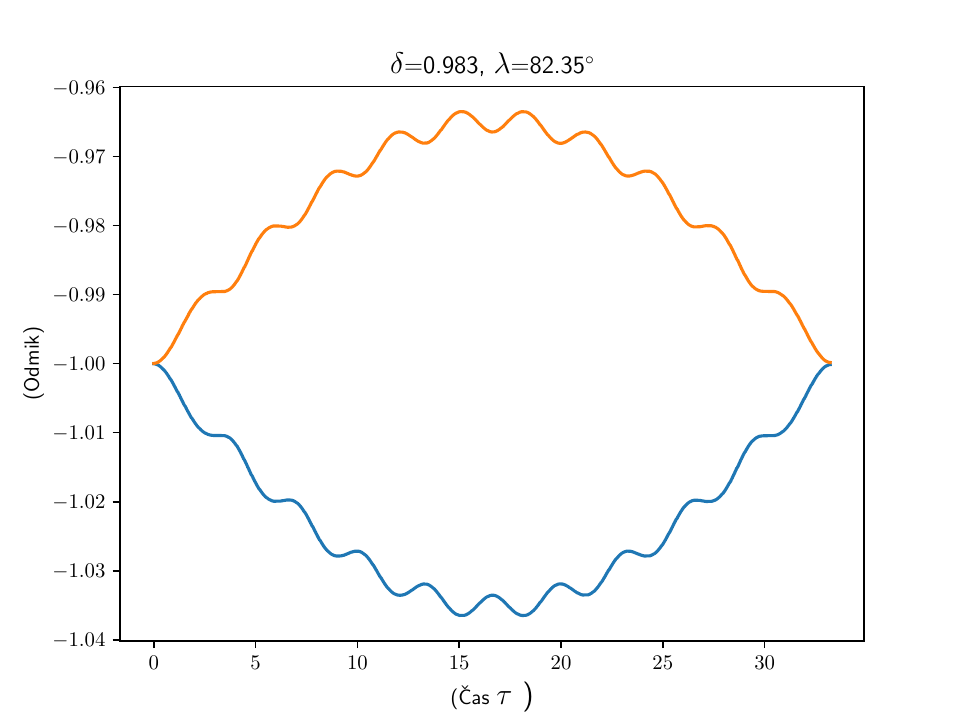}
\caption{Extreme longitudinal deviations from the rest position for
$\lambda=7.91^\circ,67.92^\circ,80.58^\circ$}\label{fig:ena}
\end{figure}

\begin{figure}[H]
\includegraphics[width=0.32\textwidth]{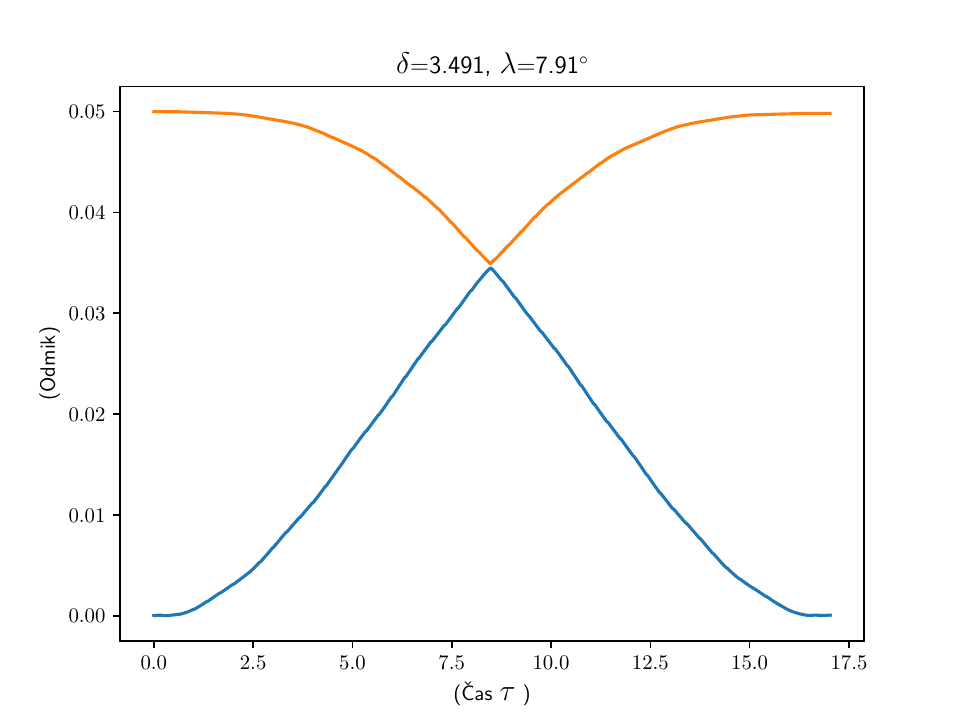}
\includegraphics[width=0.32\textwidth]{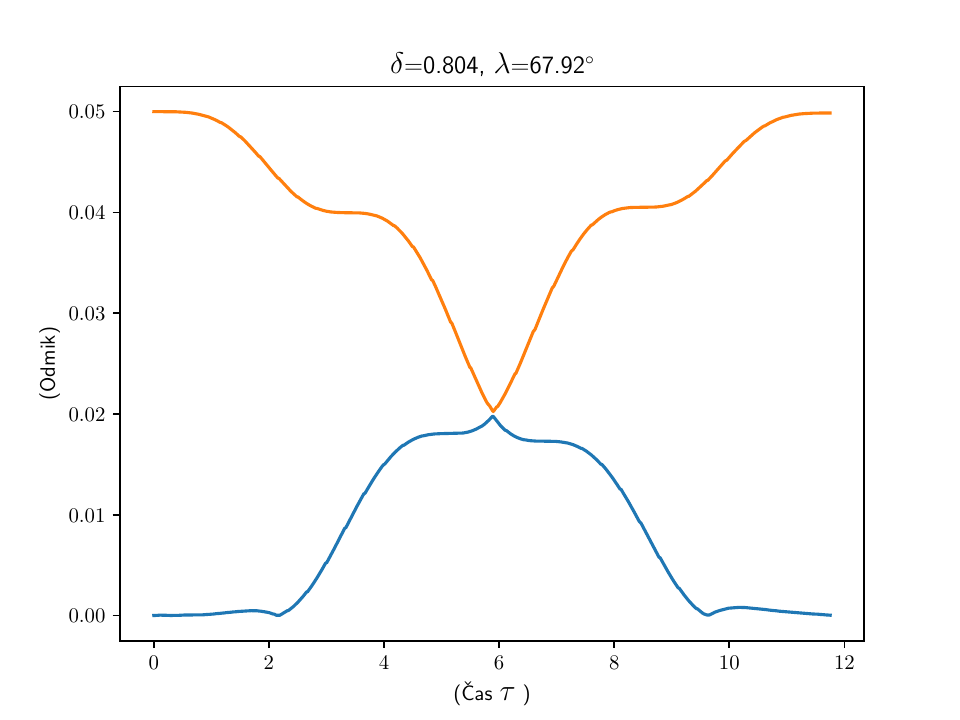}
\includegraphics[width=0.32\textwidth]{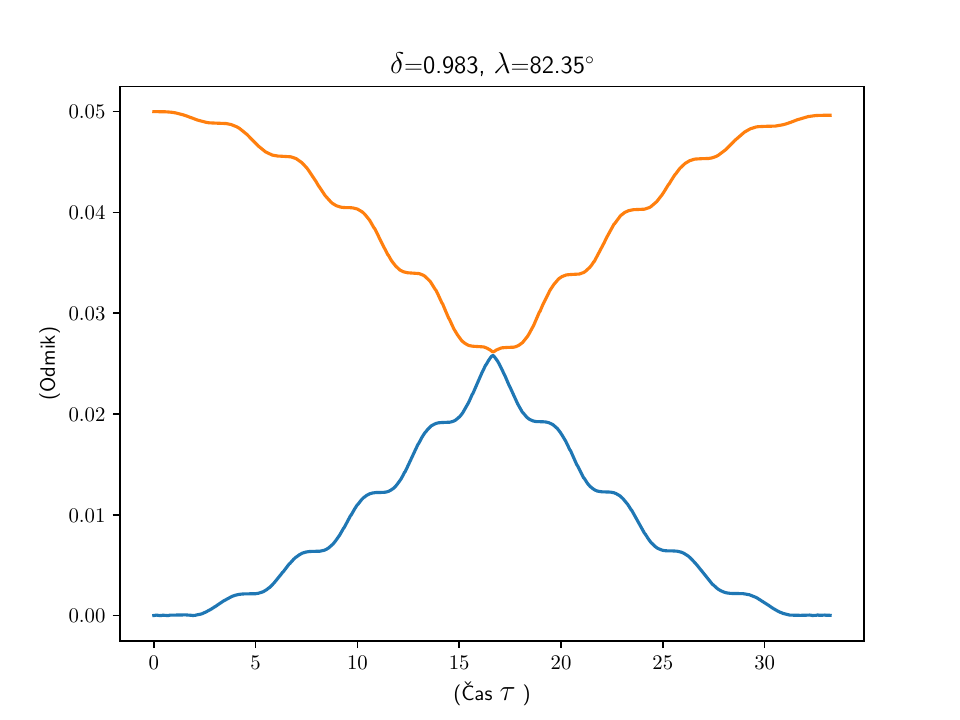}
\caption{Extreme deviations from the vertical
for $\lambda=7.91^\circ,67.92^\circ,82.35^\circ$}\label{fig:enn}
\end{figure}

In Figures \ref{fig:ena} and \ref{fig:enn}, periodically varying patterns are shown at three latitudes marked 
in Figure \ref{fig:ena} as extreme deviations from the rest position in the longitudinal direction 
and in figure \ref{fig:enn} as extreme deviation of the pendulum from the vertical.

\begin{figure}[H]
\begin{center}
\includegraphics[width=0.8\textwidth]{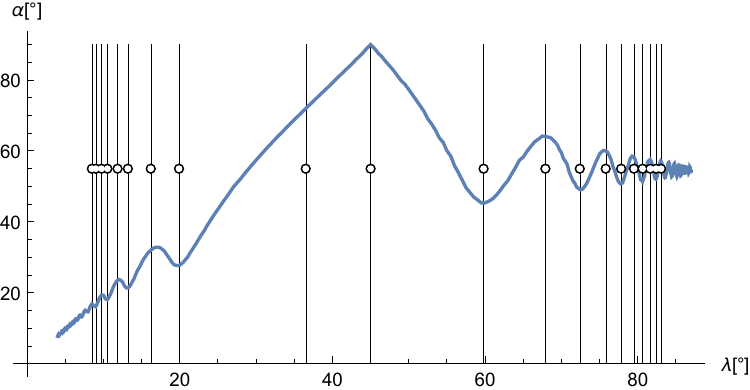}
\caption{The angle $\alpha$ on latitudes}\label{fig:alpha}
\end{center}
\end{figure}

The compound oscillation shown in Figure \ref{fig:ena} has 10 peaks at the resonance latitude 
$\lambda=7.91^\circ$, 3 peaks at $\lambda=67.92^\circ$, and 10 peaks at $\lambda=82.35^\circ$. 
The latitudes at which the pendulum oscillation with critical stiffness is periodic are listed in 
the following table along with the number of peaks of the superimposed oscillation. 
We will refer to the number of peaks of the superimposed oscillation as the order of the resonance latitude. 
We observe that the order increases as we move towards the poles or the equator, 
while the resonant latitudes become more and more dense.
\begin{center}
         \begin{tabular}{|r c|r c|r c|r c|r c|}
         \hline
         \multicolumn{10}{|c|}{Resonant latitudes}\\
         \multicolumn{10}{|c|}{$\lambda$ in degrees and their order}\\
         \hline
         7.91 & 10 & 8.43 & 9 & 8.95 & 8 & 9.72 & 7 & 10.49 & 6\\
         11.85 & 5 & 13.20 & 4 & 16.16 & 3 & 19.88 & 2 & 36.50 & 1\\
         45.00 & 1 & 60.05 & 2 & 67.92 & 3 & 72.53 & 4 & 75.61 & 5\\
         77.75 & 6 & 79.35 & 7 & 80.58 & 8 & 81.55 & 9 & 82.35 & 10\\
         \hline
         \end{tabular}
\end{center}
\noindent
Figure \ref{fig:alpha} shows the angle $\alpha$ between the $z$ axis and the normal of the trajectory
plane of the bob at different latitudes with critical string stiffness at time $\tau$. 
The pattern of the angle $\alpha$ is similar to that of the eccentricity of an ellipse and the 
radius of a circle, as shown in Figure \ref{fig:radius}.
\newpage
\section{Dependence of $\tau$ and $\delta$ on resonance latitudes}
The function expressing the dependence between the order of resonance and its latitude will be denoted by $\lambda_n = X(n)$ where negative numbers correspond to 
resonance latitudes below the $45^\circ$-th parallel and 
the order on the $45^\circ$-th parallel is $0$.
\begin{figure}[H]
\begin{center}
\includegraphics[width=0.8\textwidth]{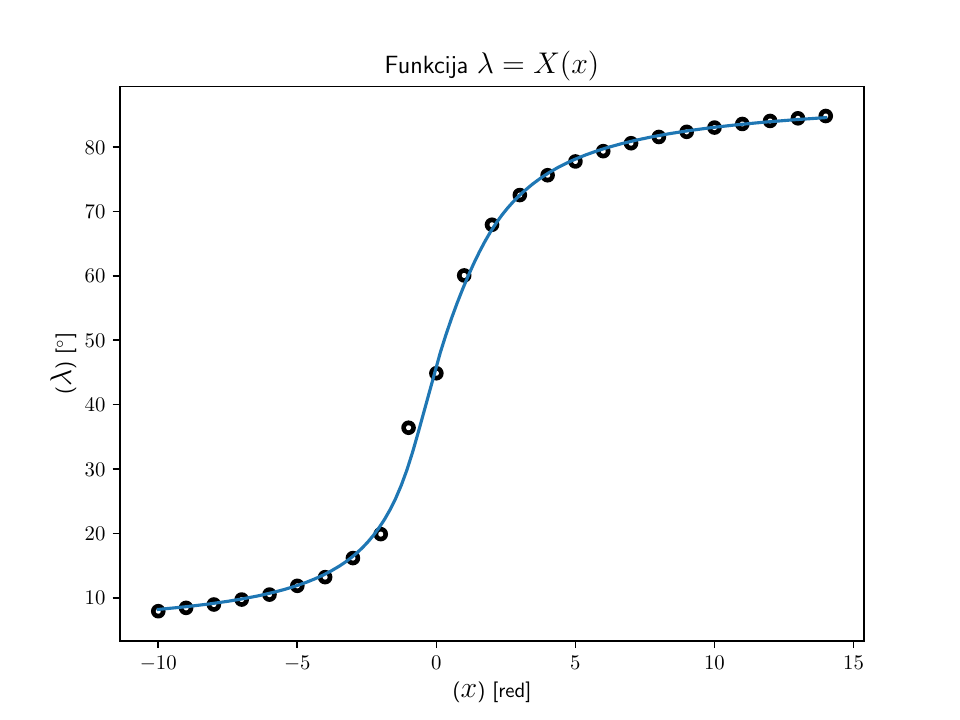}
\caption{Latitude as function of $x$}\label{fig:fx}
\end{center}
\end{figure}

The dependence function has a sigmoidal shape, as shown in Figure \ref{fig:fx}.
We interpolated the function to obtain a continuous graph, which closely resembles the graph of:
\begin{eqnarray}\label{equ:lambda}
\lambda = \frac{90}{\pi}\left(\frac{\pi}{2} + \arctan\frac{x}{2}\right).
\end{eqnarray}
However, the graph on the Figure \eqref{fig:fx} is not symmetric with respect to the point $(0, 45)$, which indicates that
the dynamics below the $45^\circ$-th parallel are different from those above this parallel.
The independent continuous variable is denoted by $x$, and the function by $\lambda = X(x)$.

In the following, we will analyze the dependencies of $\tau(x)$ and $\delta(x)$.

\newpage
\begin{itemize}

\item Time $\tau(x)$

\begin{figure}[H]
\includegraphics[width=0.495\textwidth]{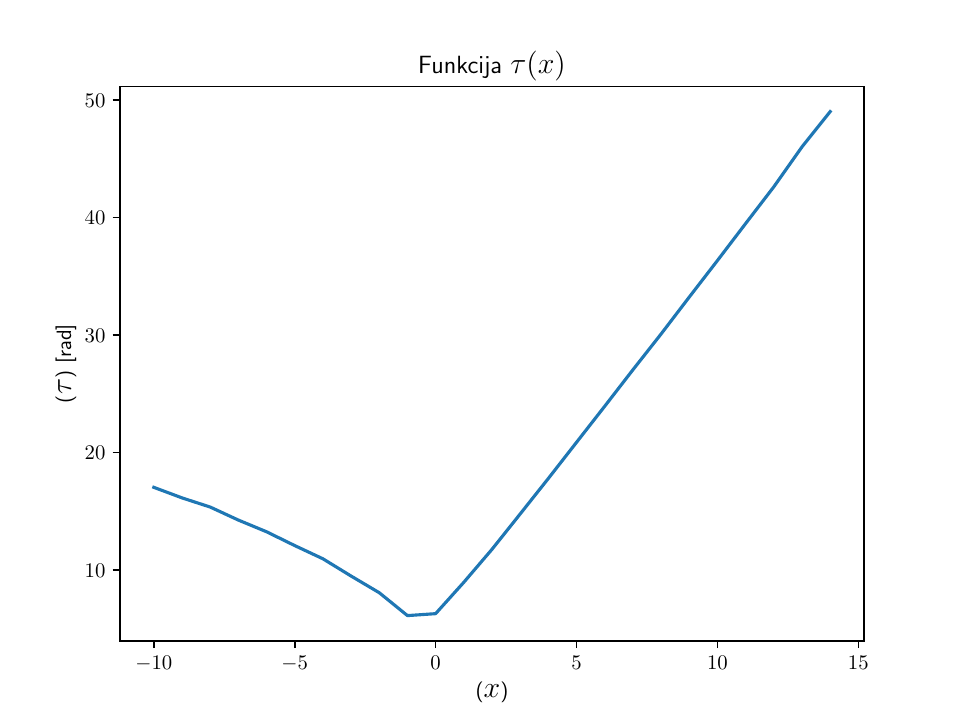}
\includegraphics[width=0.495\textwidth]{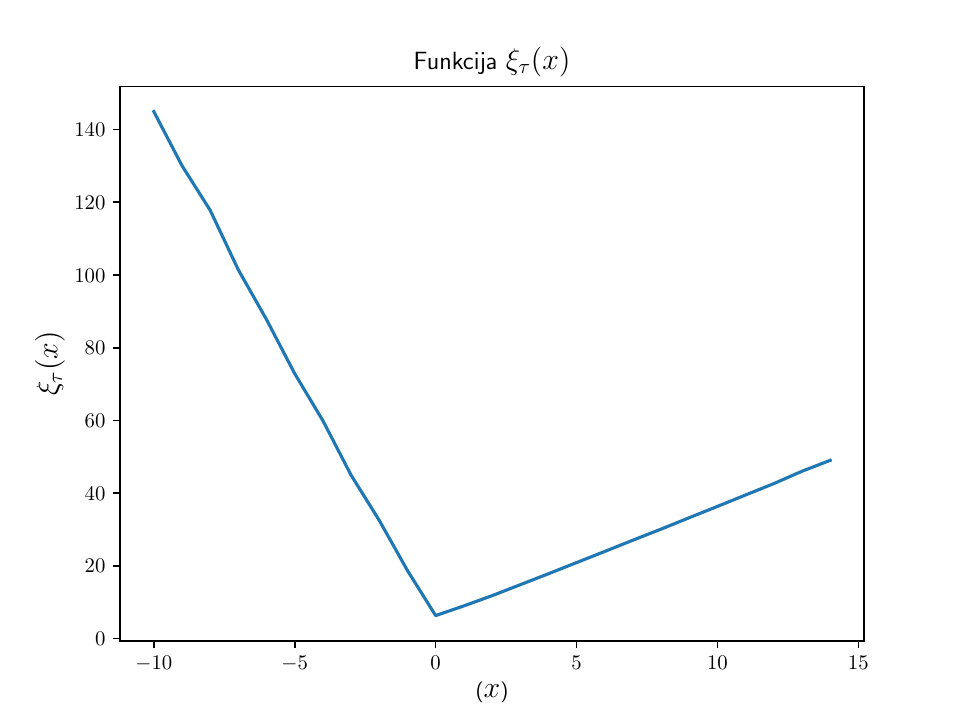}
\caption{Time $\tau(x)$ and function $\xi_\tau(x)$}
\end{figure}
\noindent
The function $\xi_\tau(x)$ with graph on the right is defined as: 
\begin{eqnarray*}
   \xi_\tau(x) = \begin{cases}
             \tau(x)^2 &\quad\text{for}\quad x <0\\
             \tau(x) &\quad\text{for} \quad  x\ge0\\
          \end{cases}.
\end{eqnarray*}
\item Critical values of the coefficient  $\delta(x)$
\begin{figure}[H]
\includegraphics[width=0.495\textwidth]{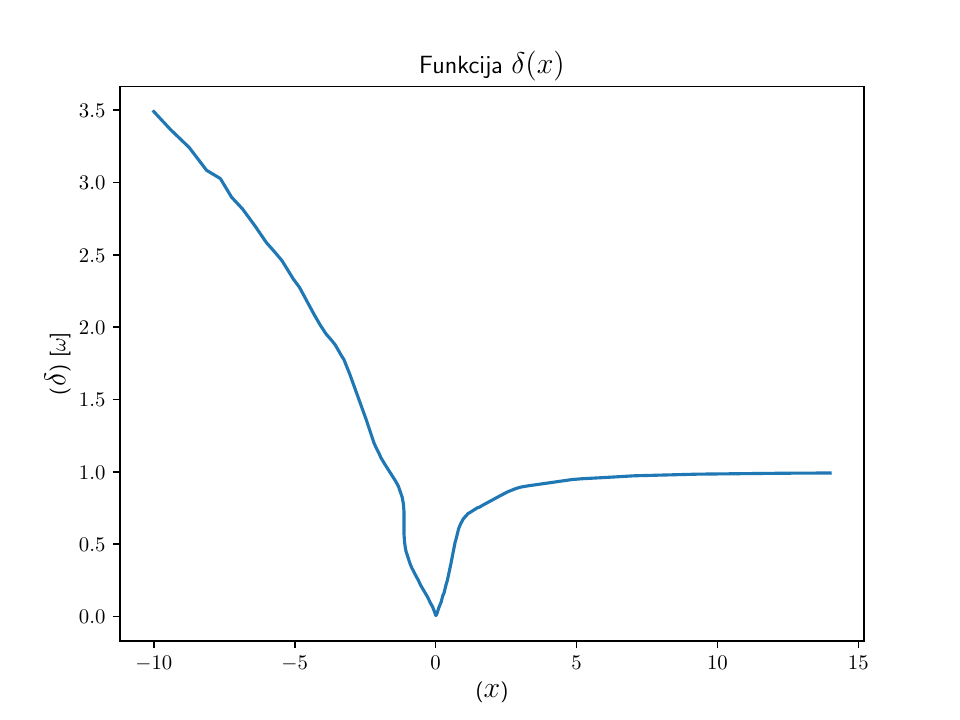}
\includegraphics[width=0.495\textwidth]{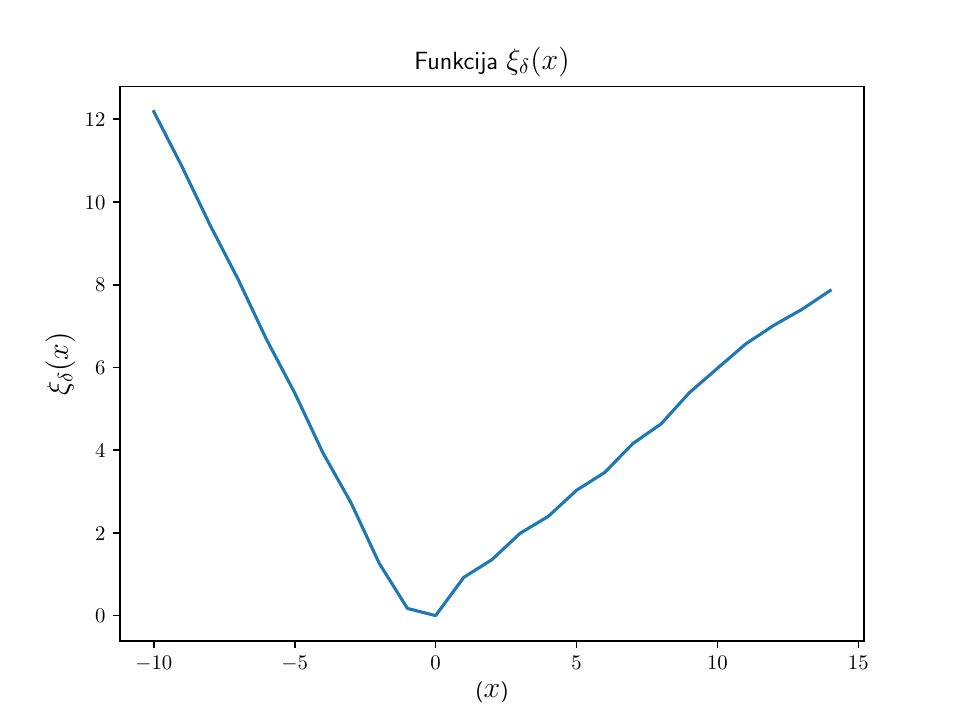}
\caption{Critical values $\delta(x)$ and function $\xi_\delta(x)$}
\end{figure}
\noindent
The function $\xi_\delta(x)$ with graph on the right is defined as: 
\begin{eqnarray*}
\xi_\delta(x) = \begin{cases}
             \delta(x)^2 &\quad\text{for}\quad x<0\\
             \delta(x)/\sqrt{1-\delta(x)^2} &\quad\text{for}\quad x\ge0\\
          \end{cases}. 
\end{eqnarray*}
\end{itemize}
\newpage
\section{Something to think about}
\begin{figure}[H]
\begin{center}
\includegraphics[width=0.8\textwidth]{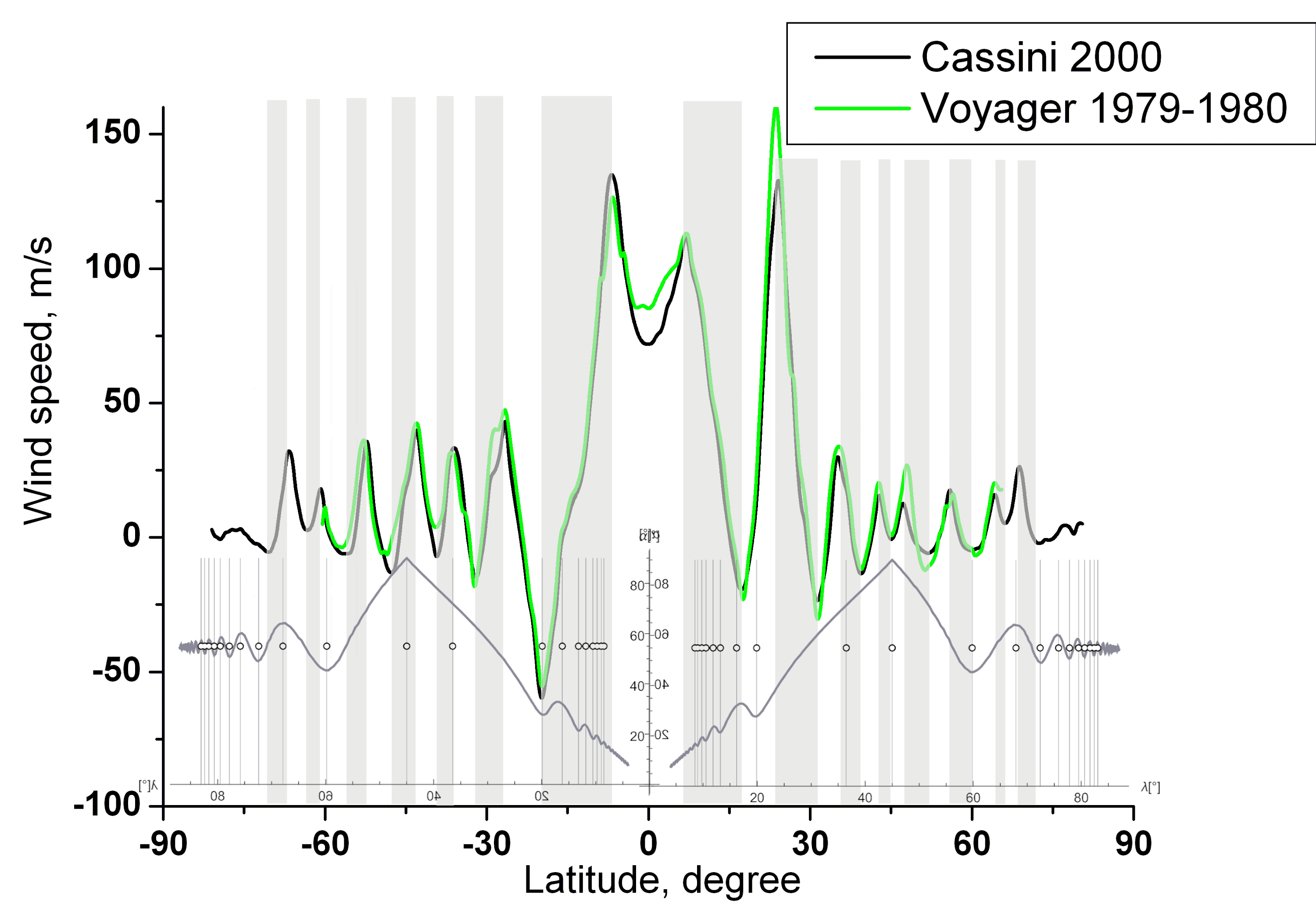}
\caption{Wind speeds on Jupiter}\label{fig:beta}
\end{center}
\end{figure}
\begin{figure}[H]
\includegraphics[width=\textwidth]{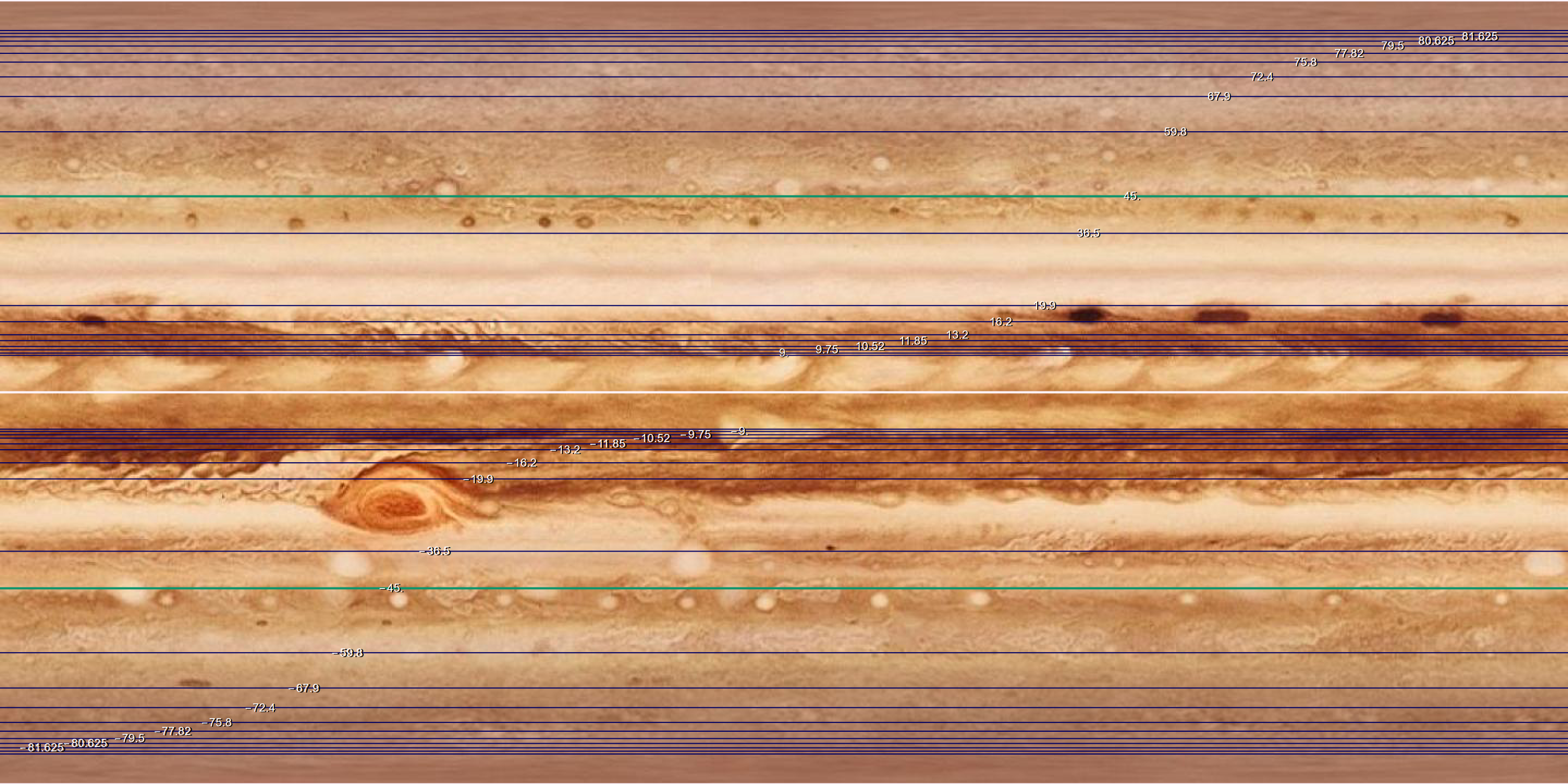}
\caption{Critical latitudes on Jupiter}\label{fig:gamma}
\end{figure}

Figure \ref{fig:beta} shows the superimposed images of Figure \ref{fig:alpha} 
and the wind speeds at the latitudes of the planet Jupiter, 
highlighting the correlation between resonant latitudes and stripes on the surface of the planet.
In Figure \ref{fig:gamma}, the resonant latitudes are marked on the image of Jupiter's surface.

\end{document}